\documentclass[conference]{IEEEtran}

\usepackage{url}
\usepackage[sort]{cite}
\usepackage[usenames, dvipsnames]{color}
\usepackage[nolist,nohyperlinks]{acronym}
\usepackage[table,xcdraw]{xcolor}
\usepackage{tabularx}
\usepackage[pdftex]{graphicx}
\graphicspath{{images/}}
\DeclareGraphicsExtensions{.pdf,.jpeg,.png}
\usepackage{array}
\usepackage{caption}
\usepackage{subcaption}

\makeatletter
\def\ps@IEEEtitlepagestyle{%
  \def\@oddfoot{\mycopyrightnotice}%
  \def\@evenfoot{}%
}
\def\mycopyrightnotice{%
    \begin{minipage}{\textwidth}
        \centering\tiny{DOI 10.1109/NFV-SDN.2017.8169876 \copyright 2017 IEEE.  Personal use of this material is permitted.  Permission from IEEE must be obtained for all other uses, in any current or future media, including reprinting/republishing this material for advertising or promotional purposes, creating new collective works, for resale or redistribution to servers or lists, or reuse of any copyrighted component of this work in other works.}
    \end{minipage}
  \gdef\mycopyrightnotice{}
}

\begin{document}

\title{Isolating SDN Control Traffic with Layer-2 Slicing in 6TiSCH Industrial IoT Networks}

\author{
\IEEEauthorblockN{
Michael Baddeley\IEEEauthorrefmark{1}, 
Reza Nejabati\IEEEauthorrefmark{1}, 
George Oikonomou\IEEEauthorrefmark{1},
Sedat Gormus\IEEEauthorrefmark{2}, \\
Mahesh Sooriyabandara\IEEEauthorrefmark{3},
Dimitra Simeonidou\IEEEauthorrefmark{1}}

\IEEEauthorblockA{\IEEEauthorrefmark{1}High Performance Networks Group, University of Bristol, Bristol, United Kingdom, \\ Email: \{m.baddeley, reza.nejabati, g.oikonomou, dimitra.simeonidou\}@bristol.ac.uk}
\IEEEauthorblockA{\IEEEauthorrefmark{2}Karadeniz Technical University, Email: sedatgormus@ktu.edu.tr }
\IEEEauthorblockA{\IEEEauthorrefmark{3}Toshiba Research Europe Ltd., Bristol, United Kingdom, Email: mahesh@toshiba-trel.com }
}
\maketitle

\begin{abstract}
Recent standardization efforts in IEEE 802.15.4-2015 Time Scheduled Channel Hopping (TSCH) and the IETF 6TiSCH Working Group (WG), aim to provide deterministic communications and efficient allocation of resources across constrained Internet of Things (IoT) networks, particularly in Industrial IoT (IIoT) scenarios. Within 6TiSCH, Software Defined Networking (SDN) has been identified as means of providing centralized control in a number of key situations. However, implementing a centralized SDN architecture in a Low Power and Lossy Network (LLN) faces considerable challenges: not only is controller traffic subject to jitter due to unreliable links and network contention, but the overhead generated by SDN can severely affect the performance of other traffic. This paper proposes using 6TiSCH tracks, a Layer-2 slicing mechanism for creating dedicated forwarding paths across TSCH networks, in order to isolate the SDN control overhead. Not only does this prevent control traffic from affecting the performance of other data flows, but the properties of 6TiSCH tracks allows deterministic, low-latency SDN controller communication. Using our own lightweight SDN implementation for Contiki OS, we firstly demonstrate the effect of SDN control traffic on application data flows across a 6TiSCH network. We then show that by slicing the network through the allocation of dedicated resources along a SDN control path, tracks provide an effective means of mitigating the cost of SDN control overhead in IEEE 802.15.4-2015 TSCH networks.
\end{abstract}


\section{Introduction}

\label{sec_intro}
\textbf{Context:} In recent years Software Defined Networking (SDN) has gained popularity as a means to bring scalability and programmability to network architecture. Particularly in Data Center and Optical Networks, SDN has been shown to offer a high degree of network reconfigurability, reduction in capital expenditure, and a platform for Network Function Virtualization (NFV) \cite{sdn_comprehensive_survey}. These benefits have led to a number of research efforts to apply SDN within the IEEE 802.15.4 low power mesh networking standard, which underpins many Industrial Internet of Things (IIoT) networks. However, these networks are constrained in terms of reliability, throughput, and energy. These limitations mean that the traditional, high-overhead approach of SDN cannot be mapped directly to a Low Power and Lossy Network (LLN). Specifically, the problem is twofold: SDN control messages can be prone to delay and jitter, and the burden of controller overhead can severely affect other network traffic flows.

\textbf{Challenge:} One of the core principles behind SDN is the separation of the control and data planes: centralizing the control plane at an external SDN controller. In a traditional, wired, SDN architecture, network peripherals such as routers, load-balancers, and firewalls, are reduced to dumb switches which can be programmed to perform these tasks through manipulation of their tables using a SDN protocol such as OpenFlow. This programmability is possible due to low latency, reliable links between network devices and the controller, ensuring that network decisions are made rapidly, and that controller decisions are made with an up-to-date global view of the network. 

In IEEE 802.15.4 networks the link between a controller and network device cannot be guaranteed. Controller communication needs to navigate multiple hops across a mesh, survive unreliable links, tackle low throughput, and compete with other traffic. In a centralized SDN architecture, this means that nodes cannot be assured of successful and timely decisions from the controller, with no guarantee on delay and packet loss. 

\textbf{Approach:} However, the recent standardization effort from the IETF 6TiSCH Working Group (WG) \cite{6tisch_ietf_architecture} aims to enable reliable IPv6 communications on top of IEEE 802.15.4-2015 Time Scheduled Channel Hopping (TSCH) networks. This is particularly relevant in industrial process control, automation, and monitoring applications: where failures or loss of communications can jeopardize safety processes, or have knock-on effects on processes down-the-line. The frequency and time diversity achieved from the TSCH schedule gives greater protection against a lossy network environment, and by specifying how that schedule is managed, 6TiSCH provides a platform to deliver Quality of Service (QoS) guarantees to network traffic. Additionally, 6TiSCH introduces the concept of \textit{tracks}: a Layer-2 mesh-under forwarding mechanism \cite{6lowpan_routing_rfc} which allows low-latency paths across the network. Tracks are created by slicing the TSCH schedule, and reserving buffer resources, at each hop along a route between a source and destination node. Each cell represents an atomic unit of bandwidth, and allocated cells are dedicated to that particular track. Since each cell is scheduled at a known time, tracks are essentially a deterministic slice of the network bandwidth, with a guaranteed minimum bounded latency \cite{6tisch_ietf_architecture}. 

\textbf{Contribution:} We aim to exploit the deterministic and low latency properties of 6TiSCH tracks to slice the TSCH schedule for the SDN control path, and provide a reliable link between constrained SDN nodes and a controller in a IEEE 802.15.4 low power wireless network. By using tracks to overcome the control link unreliability, the challenge of implementing a centralized SDN architecture within a LLN becomes more feasible, which in turn allows greater scalability and re-programmability in IoT sensor networks.

\textbf{Outline:} The rest of this paper is organized as follows. In Section \ref{sec_background} we present the problem of introducing a high-overhead SDN architecture given the constraints of a low power mesh network. We then introduce the 6TiSCH architecture, giving an overview of the terminology. Section \ref{sec_motivation} presents the main contribution of this paper, discussing how Layer-2 slicing using 6TiSCH tracks can isolate SDN control traffic in a LLN, and presents an integrated SDN/6TiSCH stack with track allocation for SDN controller links. Finally, we evaluate results of simulations in Section \ref{sec_results}, and draw conclusions in Section \ref{sec_conclusion}.

\section{Background and Related Works}
\label{sec_background}

\subsection{The Difficulties of SDN in Low Power and Lossy Networks}
\label{sec_sdn_background}

Over the last decade SDN has generated considerable academic and industrial interest, and has been successfully applied to areas such as data-center and optical networks \cite{sdn_comprehensive_survey}. However, the concept faces difficulties when trying to apply it within a wireless medium, where issues such as channel contention, interference, mobility, and topology management need to be addressed \cite{sdwn_opportunities_challenges}. Low Power and Lossy Networks face an even greater set of challenges, as they are typically made up of embedded devices where nodes are constrained in terms of energy, processing, and memory. The protocols governing these networks employ a variety of techniques in order to manage these limitations, and the IEEE 802.15.4 stack has allowed the IoT to be extended to even the smallest of devices. Yet the high-overhead approach of traditional SDN networks, where devices can enjoy large flowtables and quick responses from the controller, isn't possible within a constrained LLN environment. This ultimately forces a rethink of many of the assumptions traditionally held in SDN architecture. We attempt to lay out these assumptions, and show how they are challenged by the realities within a LLN.

\textbf{Low-Latency Controller Communication:} In traditional SDN architecture, low-latency links between SDN devices and the controller allow rapid state and configuration changes. This ability, to be able to quickly convey network information and decision making between the control and data planes, is vital within a SDN network. The constraints of IEEE 802.15.4 mesh networks mean that traffic faces maximum transfer rates of 250kbit/s, 127B frame sizes before fragmentation (depending on the 6LoWPAN \cite{6lowpan_rfc}), and a multi-hop mesh topology where processing and link overhead is incurred at each hop.

\textbf{Reliable Links:} In placing an external SDN controller, there is an inherent assumption that network devices will be able to reliably communicate with that controller. However, the lossy medium in LLNs means that this reliability cannot be guaranteed - a direct consequence of the low power requirements of many IoT sensor networks. The problem becomes more acute when traversing multiple hops, as an element of uncertainty is introduced at each link. Depending on the buffers at each node, and the losses incurred at each hop, this can lead to a large amount of jitter.

\textbf{Dedicated Control Channel:} In addition to the physical layer issues such as fading and interference, SDN control messages must compete with both topology and synchronization protocols, as well as regular application traffic. This contention can lead to retransmissions and lost packets. Given the energy limitations of the devices, this can be undesirable.

\textbf{Large Flowtables:} In a traditional wired SDN implementations device hardware will support multiple, large flowtables. These tables are often able to hold thousands or tens of thousands of entries, allowing SDN devices to be configured for many different types of data flows and perform a number of network functions. LLNs are constrained not only by their environment, but also by the devices themselves. In order to reduce costs and energy expenditure,  devices are typically limited in their processing power and only support a few kB of memory, meaning that nodes cannot store more than a handful of flowtable entries.

\textbf{Real-Time Network State:} A core principle of SDN is the assumption that the network will be able to maintain an abstraction of the distributed network state, mirroring the changes made in the physical network in near real-time. The problems of latency and reliability in a LLN mean that there is a great deal of uncertainty within the network, and this uncertainty is reflected in the abstraction model of the network state. Without this accuracy there can only be limited confidence attributed to controller decisions.


\subsection{6TiSCH Architecture and Terminology}
\label{sec_6tisch_background}

Recent work from the IETF 6TiSCH standardization Working Group (WG) aims to incorporate concepts from the IETF Deterministic Networking \cite{ietf_detnet} and SDN \cite{ietf_sdn_rfc} WGs, enabling deterministic, centralized scheduling of communication across IEEE 802.15.4-2015 TSCH networks. The charter tasks the WG with producing specifications for how nodes communicate the TSCH schedule between themselves, default scheduling functions for dynamic scheduling of timeslots for IP traffic, and the detailing of an interface to allow deterministic routes across the 6TiSCH LLN through the manipulation of 6TiSCH scheduling and forwarding mechanisms. This section attempts to introduce the main concepts of 6TiSCH and its associated terminology, as well as explore how the SDN concept is imagined within the architecture.

\begin{figure}[ht]
\centering
  \includegraphics[width=1.0\columnwidth]{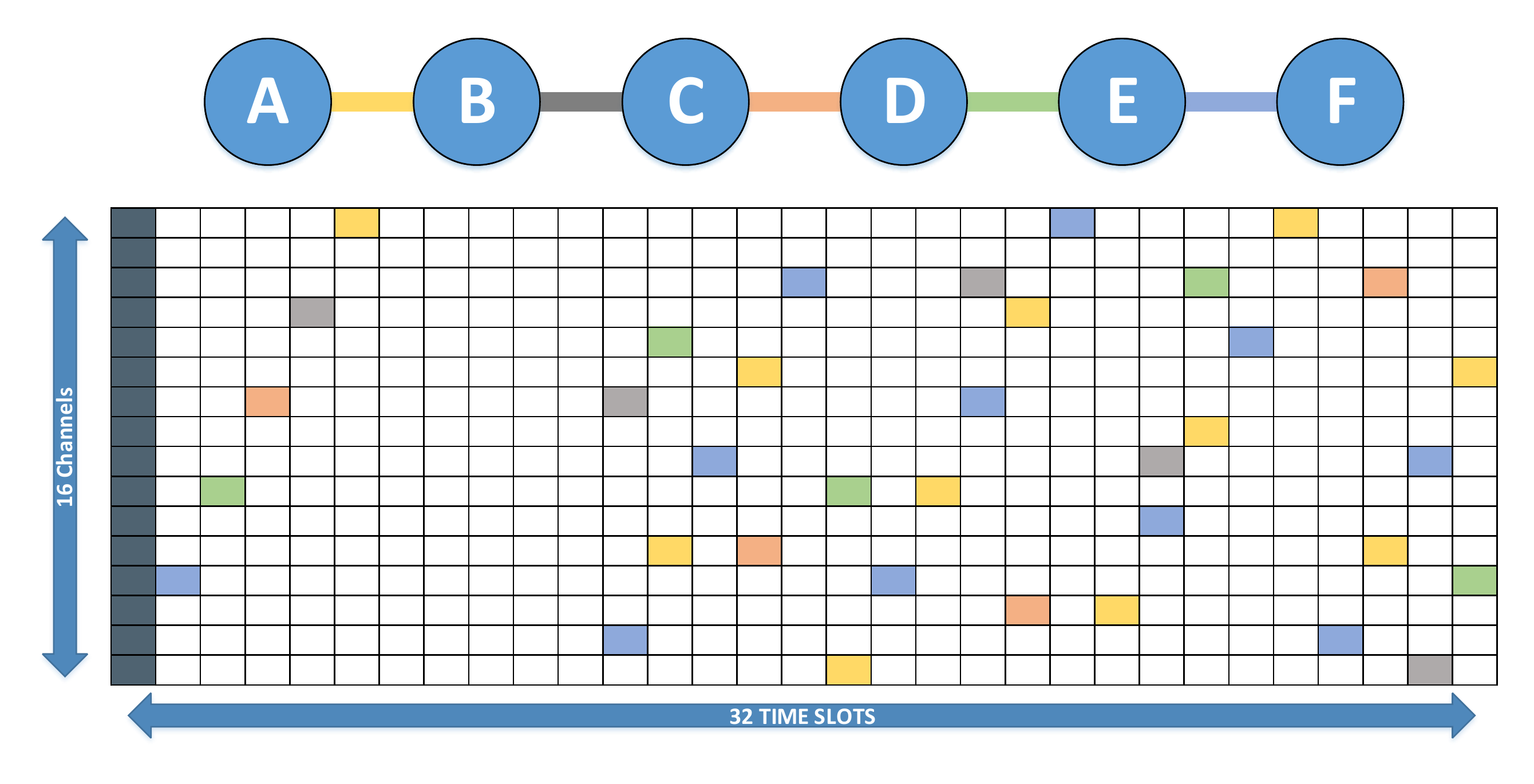}
  \caption{Allocation of TSCH resources in a 5 hop network, with dedicated SDN control slices.}
  \label{fig:track_schedule}
\end{figure}

\textbf{IEEE 802.15.4-2015 TSCH:} 
Time Scheduled Channel Hopping (TSCH) is a deterministic MAC layer that introduces channel hopping across a Time Division Multiple Access (TDMA) schedule. This frequency diversity allows for greater reliability and energy efficiency than CSMA-based MAC layers, particularly in environments where there is a high degree of fading and interference. The TSCH schedule is sliced into repeating slotFrames [sic], consisting of timeslots spread over a number of frequency channels. Nodes periodically exchange synchronization beacons, which contain startup information about the TSCH schedule. Each timeslot can be allocated as a dedicated Transmit (Tx) or Receive (Rx) opportunity for that node, used as a "sleep" cell, or shared with other nodes in a contention-based scheme. By creating a schedule across the spectrum, interference is reduced, as it ensures that nodes within interference range aren't attempting to transmit simultaneously. In addition, nodes only need to be awake for Tx, Rx, and Shared cells, improving their energy consumption. 

\textbf{6top Sublayer:} 
The 6TiSCH Operation Sublayer (6top) allows 6TiSCH to communicate TSCH scheduling information to and from nodes within the network, and provides mechanisms for both distributed and centralized scheduling. In the distributed scenario, schedule information is communicated between nodes using the 6top Protocol (6P), transported on 6TiSCH Information Elements (IEs). In a centralized scheduling scenario, schedule information is communicated to and from a centrally located Path Computation Engine (PCE), analogous to a SDN controller, through the 6top CoAP Management Interface (COMI).

\textbf{Schedule Management:}
The 6TiSCH standard allows for a number of schedule management paradigms. These mechanisms allow for flexible maintenance of the TSCH schedule, based upon the the needs of the network. In \textit{Static Scheduling} a common fixed schedule is shared by all nodes in the network, and can be used as a network initialization mechanism. It is equivalent to Slotted ALOHA, and remains unchanged once a node joins the network. \textit{Neighbor-to-Neighbor} allows distributed scheduling functions by providing mechanisms so that matching portions of the TSCH schedule can be established by and between two neighboring nodes. As part of the charter, 6TiSCH is tasked with incorporating elements of SDN architecture: for \textit{Remote Schedule Management}, a centralized Path Coordination Engine (PCE) is able to make schedule changes based on the overall network state collected from each node. Finally, a \textit{Hop-by-Hop} scheduling mechanism is used for the distributed allocation of 6TiSCH tracks, and is the method we use in this paper. With Hop-by-Hop scheduling, a node can reserve a slice of the TSCH schedule as a dedicated Layer-2 forwarding path towards a destination, by getting 6top to allocate cells at each intermediate node.

\textbf{Routing and Forwarding:} 
6TiSCH supports three forwarding models: \textit{IPv6 Forwarding}, where each node decides on the forwarding path based on its own forwarding tables; \textit{6LoWPAN Fragment Forwarding}, where successive fragmented packets are forwarded onto the destination of the first fragment; and finally, \textit{Generalized Multi-Protocol Label Switching (G-MPLS) Track Forwarding}, which switches frames at Layer-2 based on dedicated ingress and egress cell bundles along a path.

\begin{table}[ht]
	\renewcommand{\arraystretch}{1.0}
	\caption{Summary of 6TiSCH scheduling, routing and forwarding mechanisms.}
    \label{table:6tisch_shed_rout_fwd}
	\centering
    \begin{tabular}{ |m{3cm}|l|l| }
    \hline
      	\bfseries Scheduling & \bfseries Forwarding & \bfseries Routing\\ \hline
		Static & IPv6 + 6LoWPAN Frag. & RPL \\ \hline
        Neighbor to Neighbor & IPv6 + 6LoWPAN Frag. & RPL \\ \hline
        Remote Monitoring and Schedule Management & G-MPLS Track Fwd. & PCE \\ \hline
        Hop-By-Hop & G-MPLS Track Fwd. & Reactive P2P  \\
    \hline
    \end{tabular}
\end{table}


\subsection{IETF 6TiSCH Tracks: Deterministic Layer-2 Slices}

Importantly, the concept of 6TiSCH forwarding \textit{tracks} are suggested as a mechanism for providing QoS guarantees in industrial process control, automation, and monitoring applications, and where failures or loss of communications can jeopardize safety processes, or have knock on effects on processes down-the-line. Essentially the 6TiSCH interpretation of Deterministic Paths \cite{6tisch_ietf_architecture,ietf_detnet}, tracks exhibit deterministic properties through the reservation of constrained resources (such as memory buffers), and the dedicated allocation of TSCH slotFrame cells at each intermediate node. 

Tracks are a form of Generalized Multi-Protocol Label Switching (G-MPLS), and frames are switched at Layer-2 based on the ingress cell bundle at which they were received, and forwarded to a paired transmission cell bundle. Cell bundles are groups of cells represented by a tuple consisting of \textit{\{Source MAC, Destination MAC, Track ID\}}, with the number of cells within the bundle representing the allotted bandwidth for the track. Successive bundle pairs at each intermediate node create a low-latency point-to-point path between a source and destination. As packets do not need to be delivered to Layer-3, there is less process overhead at each node. In addition, the dedicated buffer and slotFrame resources means the likelihood of retransmissions and congestion loss is reduced, as frames sent along the track don't need to compete with other traffic, reducing jitter considerably.


\section{Isolating SDN Control Traffic Overhead \\ with 6TiSCH Tracks}
\label{sec_motivation}

In this section we firstly examine how the controller overhead generated by SDN can have an adverse affect on regular network traffic flows. We introduce $\mu$SDN, our lightweight SDN implementation written for Contiki OS, in order to characterize SDN control traffic, and show the effect of different control traffic types on regular network traffic. We then propose that, in order to address the effects of SDN overhead within IEEE 802.15.4-2015 TSCH networks, 6TiSCH tracks can be utilized to provide an isolated network slice to SDN control traffic: delivering low latency SDN controller communication with minimal jitter, and minimizing disruption to the rest of the network.

\begin{figure}[ht]
\centering
	\includegraphics[width=0.6\columnwidth]{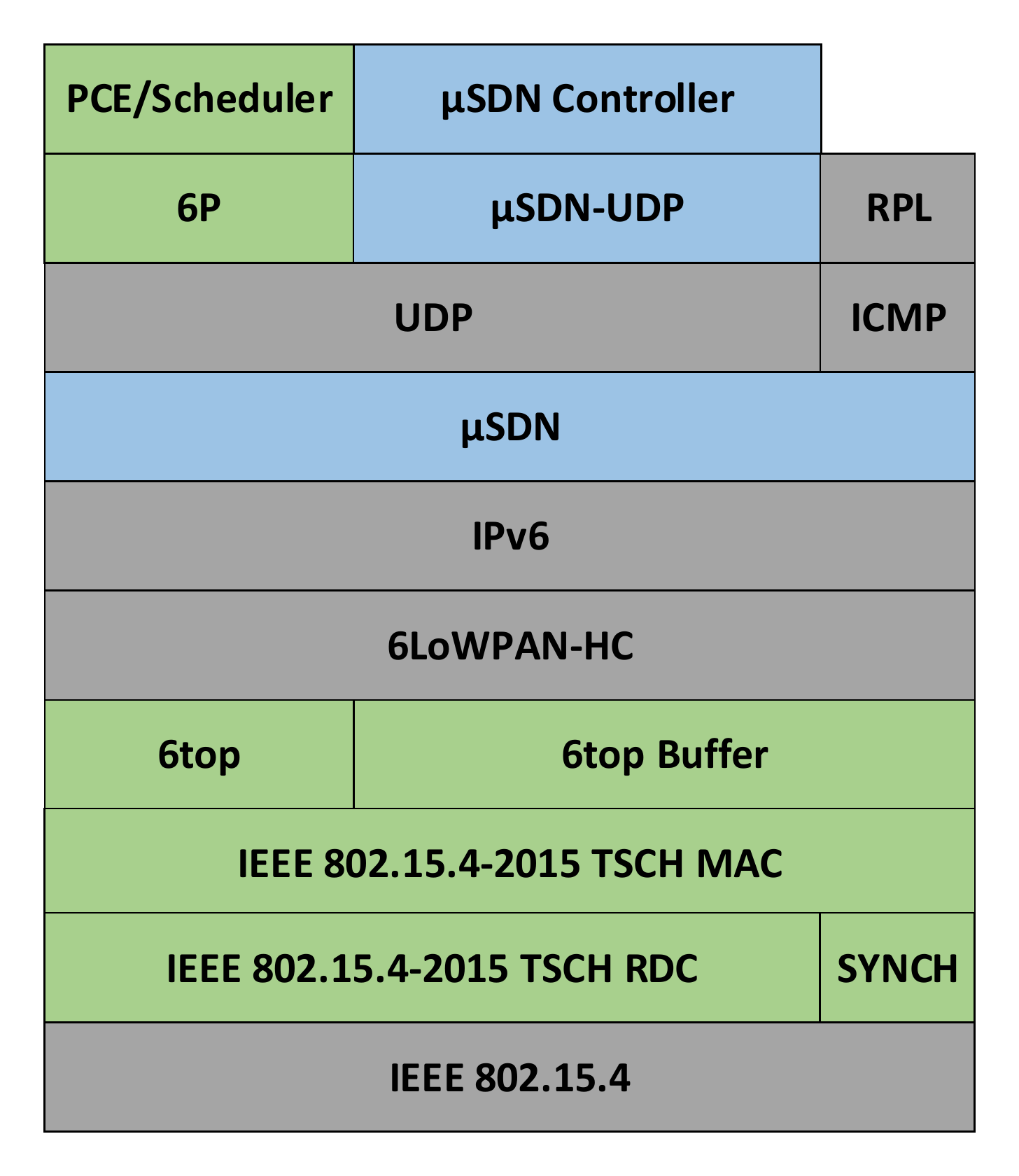}
  \caption{$\mu$SDN and 6TiSCH protocol stack.}
  \label{fig:stack_6tisch_usdn}
\end{figure}

$\mu$SDN incorporates features for minimizing controller overhead, whilst supporting a fully programmable SDN flowtable, and is integrated with the Contiki IEEE 802.15.4-2015 stack using our own 6TiSCH implementation. It has been tested in Cooja using TI's exp5438 platform with MSP430F5438 CPU and CC2420 radio. Figure \ref{fig:stack_6tisch_usdn} shows the integration of $\mu$SDN within the 6TiSCH protocol stack. 

Traffic generated with $\mu$SDN, using the $\mu$SDN protocol exhibits different behavior depending on the packet type, and the effect that SDN overhead has on regular network traffic is related to this behavior. $\mu$SDN nodes introduce two main forms of SDN control overhead. Firstly, a Node State Update (NSU )message is periodically sent from a node to the SDN controller, and carries information about the state of that node, such as energy and buffer congestion, as well as its local network state. These messages are used to update the controller on the state of the overall network, and are sent on a timer that is configured as the node joins the SDN network. They will continue to be sent until the controller reconfigures this timer. Secondly, Flowtable Query (FTQ) messages are sent to the controller in response to a flowtable miss (i.e. the SDN process checks the flowtable for instructions on how to handle a packet but is unable to find a matching entry). These packets contain information about the packet that caused the miss. The timing of FTQ messages therefore depends on the number of 'new' packets seen by a node, as well as the flowtable entry lifetime period. Once a flowtable entry expires, the node has to query the controller for new instructions about what to do with that packet type; this process is analogous to the \textit{PacketIn}/\textit{PacketOut} process in OpenFlow. If a number of nodes have similar expiry times for flowtable entries this can result in bursts of queries being sent towards the controller. In addition, successive packets arriving at a node could cause multiple queries to be sent whilst the node waits for a response from the controller. 

\begin{figure}[ht]
\centering
  \includegraphics[width=0.8\columnwidth]{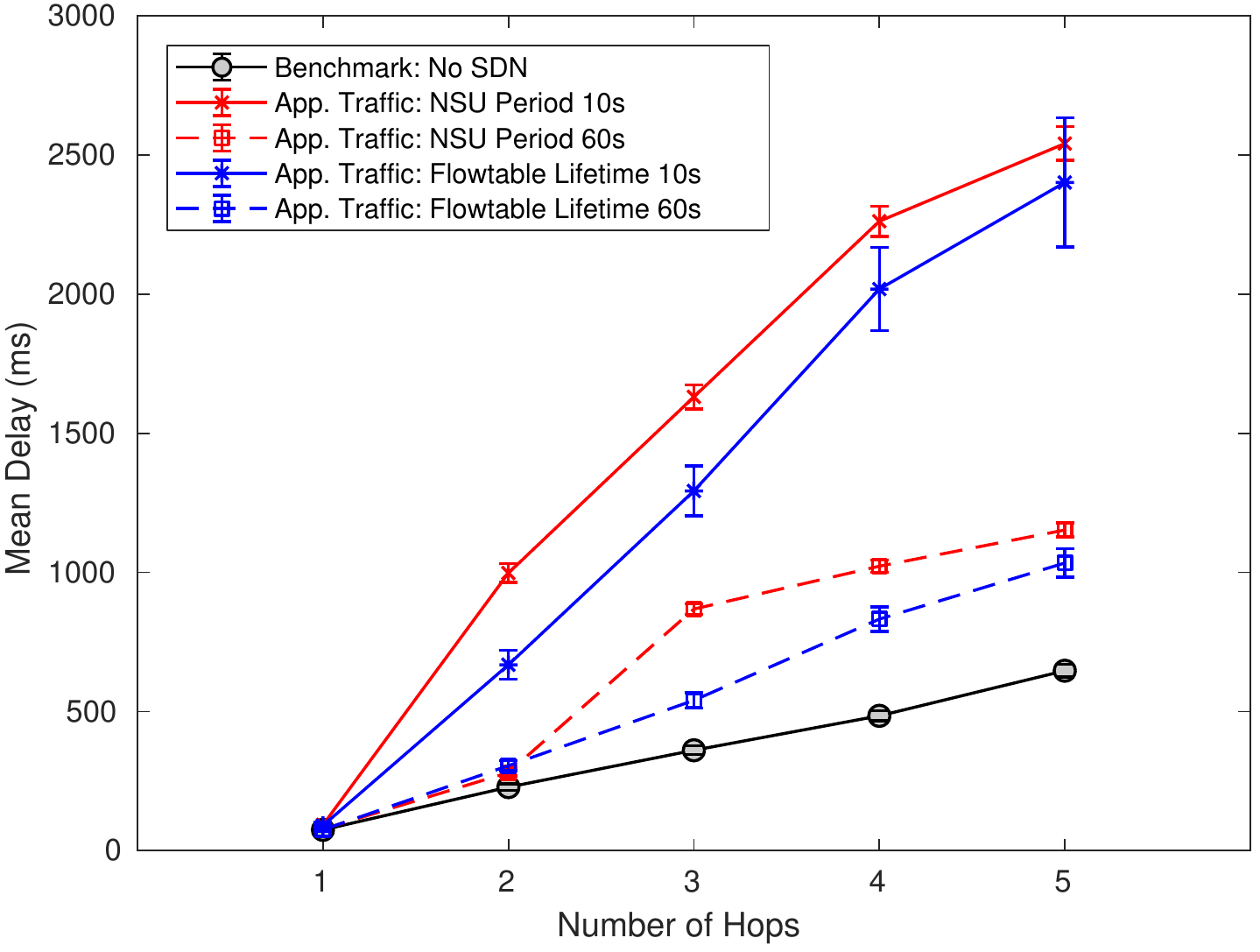}
  \caption{Effect of SDN control overhead on application-layer traffic.}
  \label{fig:nsu_1min}
\end{figure}

The SDN control traffic generated from the controller update and query processes can therefore be both \textit{periodic} as well as \textit{intermittent} and \textit{bursty}, as summarized in Table \ref{table:usdn_traffic_characterization}. However, by allocating dedicated slices for control traffic, tracks can mitigate the effects of SDN overhead on the normal operation of the network, as well as providing dedicated resources for low-latency controller communication.

\begin{table}[ht]
	\renewcommand{\arraystretch}{1.0}
	\caption{Traffic Categorization of $\mu$SDN Packet Types} 
    \label{table:usdn_traffic_characterization}
	\centering
    \begin{tabular}{ |l|l|l| }
    \hline
      	\bfseries Packet Type & \bfseries Behavior & \bfseries Effect\\ \hline
		NSU (Update) & Periodic & Increase Delay \\ \hline
        FTQ (Query) & Bursty & Increase Jitter and Packet Loss \\ 
    \hline
    \end{tabular}
\end{table}

\section{Results and Performance Evaluation}
\label{sec_results}

We evaluate the performance of both SDN control traffic and application-layer data traffic when allocating IETF 6TiSCH tracks using the Cooja network simulator for Contiki. As well as providing a network simulation, Cooja emulates the physical hardware, allowing the 6TiSCH and $\mu$SDN stacks to run on top of Contiki OS as they would in real-world scenarios. We used EXP5438 emulated motes to run the simulations, which provide 32KB of ROM and 16KB of RAM memory. Due to the memory limitations of the hardware it was not possible to run simulations with more than a handful of motes, as the 6TiSCH/$\mu$SDN code requires a large amount of RAM for buffers. We therefore throttled the available traffic bandwidth traffic in a 5-hop network, in order to emulate the increased traffic rate seen within a larger mesh, as seen in Figure \ref{fig:track_schedule}.

Section \ref{sec_motivation} demonstrated the effect of $\mu$SDN control traffic in a IEEE 802.15.4-2015 network. Specifically, we characterized the \textit{periodic} and \textit{intermittent bursty} traffic behavior of NSU and FTQ messages, used to update the controller about node state and query the controller as the result of flowtable misses. Using this traffic characterization, we performed simulations evaluating the latency and jitter of application-layer traffic: firstly in a regular \textit{6TiSCH scenario} using a distributed scheduler which allocates timeslots for optimal latency; and secondly in a 6TiSCH \textit{track scenario}, where each node has allocated a track towards the controller. In both cases UDP application traffic was sent using the resources allocated through the scheduler. However, in the track scenario, control traffic was sent on the dedicated track slice allocated for each node. The objective of these simulations was to analyze the capability of tracks as Layer-2 network slices for SDN control route isolation, and demonstrate their potential to reduce contention with regular network flows. All parameters used for the simulations parameters are summarized in Table \ref{table:results_params}.

\begin{table}[ht]
	\renewcommand{\arraystretch}{1.0}
	\caption{Simulation Parameters}
    \label{table:results_params}
	\centering
    \begin{tabular}{ |l|l| }
    \hline
      	\bfseries Simulation Parameter & \bfseries Value \\ \hline
        Transmission Range & 100m \\ \hline
        Radio Medium & UDGM (Distance Loss) \\ \hline
        Link Quality & 90\% \\ \hline
        App. Data (UDP) Interval & 5-10s \\ \hline
        SDN Update Rate (Periodic) & 10s \\ \hline
        SDN Query Rate (Intermittent) & 60s \\ \hline
        RPL Route Lifetime & 10min \\ \hline
        RPL Default Route & $\infty$ \\ \hline
        TSCH Shared Slots & 4 \\ 
    \hline
    \end{tabular}
\end{table}

\begin{figure}[ht]
\centering
  \includegraphics[width=0.8\columnwidth]{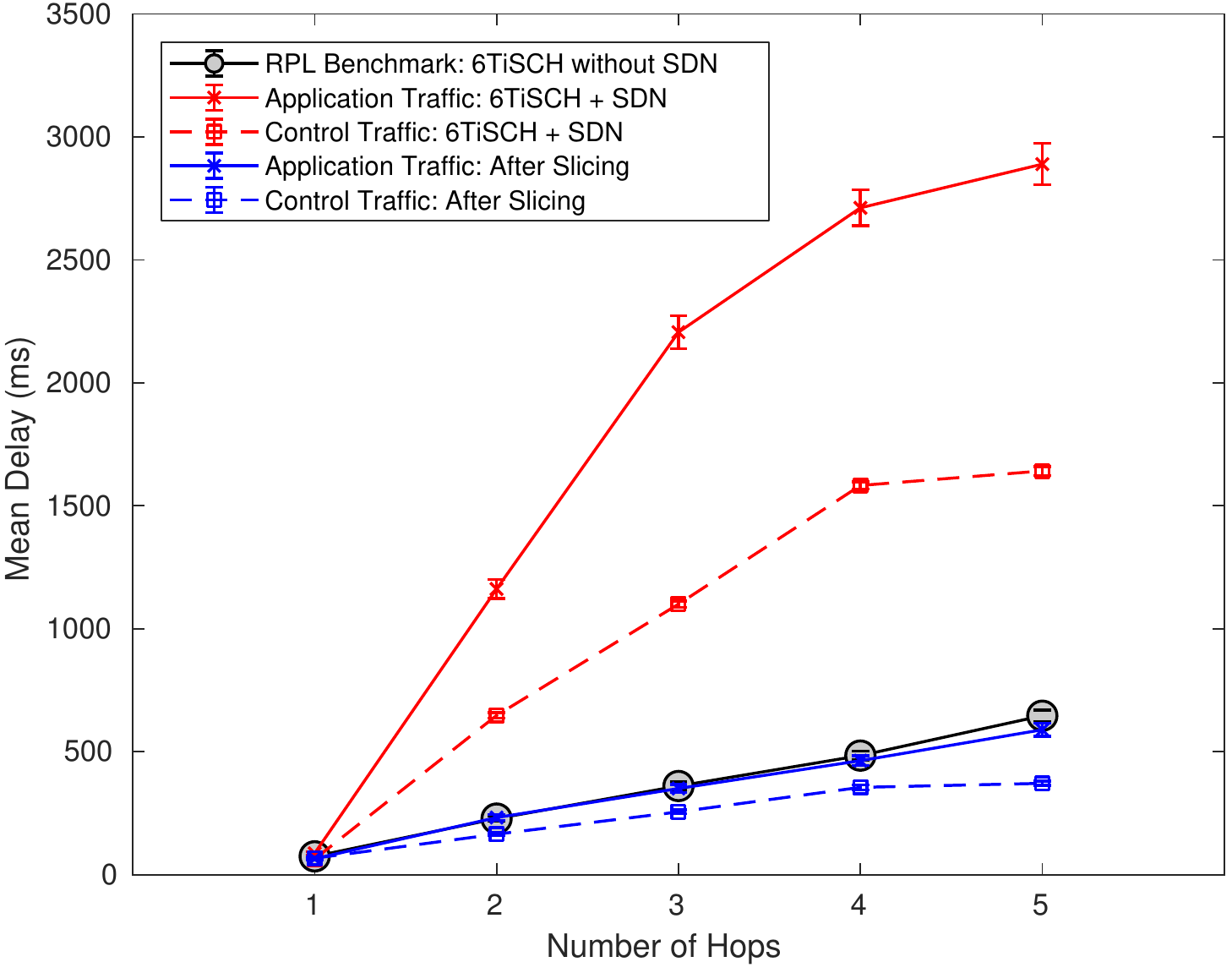}
  \caption{Delay and jitter of application and SDN traffic in a 6TiSCH SDN network with a distributed scheduler (optimizing for minimal latency), and a 6TiSCH SDN network after track slicing for SDN control flows. Results are benchmarked against a 6TiSCH scenario without SDN.}
  \label{fig:overhead_delay}
\end{figure}

Figure \ref{fig:overhead_delay} presents the results of the two simulation scenarios: a base 6TiSCH SDN network, and a 6TiSCH SDN network using tracks. We have included a \textit{NO SDN Case + RPL} benchmark to show application traffic performance in a 6TiSCH network without any SDN overhead. 

The results show latency and jitter for both application-layer traffic and SDN control traffic in each scenario. When tracks are allocated between the nodes and the controller, it is demonstrated that delay and jitter are reduced considerably: both for SDN control traffic as well as application data flows. As each track is essentially its own slice of the network, the SDN control overhead is isolated from the rest of the network, preventing interference and buffer contention. The application traffic in the track scenario performs marginally better than the \textit{NO SDN Case + RPL} benchmark. This is due to small differences in the SDN forwarding process overhead, when dealing with application traffic, compared to the RPL forwarding process overhead. In addition, the control traffic in the track scenario also performs better than the benchmark: as each track has dedicated network resources, buffer contention is eliminated for SDN control message.

\section{Conclusion}
\label{sec_conclusion}

There has been considerable interest in applying SDN concepts within IoT in order to introduce greater network programmability and provide a platform for virtualizing network functions and services. Though there has been progress towards defining and addressing the unique challenges posed by implementing SDN architecture within constrained wireless networks, one of the fundamental issues has been the inherent unreliability between between the SDN controller and nodes within the mesh. This paper has presented initial efforts to provide dedicated SDN control slices in an SDN-enabled IEEE 802.15.4-2015 network. We have demonstrated that by using IETF 6TiSCH tracks, we can create an isolated network slice for SDN control traffic; ensuring that the SDN overhead will not interfere with the performance of other network flows, whilst providing deterministic properties to SDN controller communications.

\section*{Acknowledgements}

The authors wish to acknowledge the financial support of the Engineering and Physical Sciences Research Council (EPSRC) Centre for Doctoral Training (CDT) in Communications (EP/I028153/1), as well as support from the Tubitak 115c116 project and Toshiba Research Europe Ltd. 

\bibliographystyle{IEEEtran}
\bibliography{main}

\end{document}